\documentclass[12pt]{article}

\usepackage{latexsym,amsmath,amssymb,amsthm,amsfonts,graphicx}
\usepackage{natbib}
\usepackage{epsfig,bm,authblk,url,blkarray}

\usepackage{float} 
\usepackage{booktabs} 
\usepackage{tabularx}
\usepackage{graphicx} 
\usepackage[margin=1cm]{caption} 
\usepackage{mathtools}
\usepackage{xcolor}
\usepackage{color}
\floatstyle{plain}
\usepackage{subfigure}
\usepackage{siunitx}
\usepackage{algcompatible}
\usepackage{algorithmicx}
\usepackage[noend]{algpseudocode}
\usepackage{algorithm}
\usepackage{diagbox}
\newfloat{Algorithm}{thp}{lop}
\floatname{Algorithm}{Algorithm}

\usepackage{titling}
\begin{document}

\title{Learning the distance for ABC and localized neural posterior estimation}
\date{\empty}

\author{Yuyan Wang\thanks{\textit{Department of Statistics and Data Science, National University of Singapore}.}\,\,
and 
David J. Nott\thanks{Corresponding author:  standj@nus.edu.sg. \textit{Department of Statistics and Data Science, National University of Singapore} and \textit{Institute of Operations Research and Analytics, National University of Singapore}.}}

\maketitle
\vspace{-0.7in}

\begin{abstract}
Likelihood-free inference methods can perform Bayesian inference when evaluating
the likelihood is impractical but simulating synthetic data from the model 
is feasible.  
Approximate Bayesian computation (ABC) is a well-established likelihood-free approach that constructs particle posterior approximations by evaluating the similarity between simulated and observed data using a distance function, which is used in rejection or weighting steps.
Here we extend previous work on adaptive distance learning for ABC to
misspecified time series, while also exploring applications in neural
posterior estimation using prior-data fitted networks (NPE-PFN)
with localization.  The adaptation of the distance
that we consider optimizes out-of-sample predictive performance using a 
scoring rule.  
We also establish a connection between linear pooling for forecast combination and our posterior estimation methods with randomized distances, showing that empirical estimation of pooling weights can be interpreted as another form of adaptive distance learning.
For both ABC algorithms and NPE-PFN methods with localization,
adaptive distance learning improves forecasting performance in simulated
and real examples. 

\smallskip
\noindent \textbf{Keywords:}   Bayesian inference;  forecasting;  model misspecification; simulation-based inference. 

\end{abstract}

\section{Introduction}\label{sec:Introduction}

Likelihood-free inference (LFI), also called simulation-based inference (SBI), uses model simulation to perform
Bayesian inference in models where likelihood computation is impractical, but it is possible to 
simulate synthetic data from the model.  A well-established likelihood-free approach is approximate
Bayesian computation (ABC), which requires specification of a distance function, 
with distances between synthetic and observed data being used in rejection and weighting steps
for constructing particle posterior approximations.  Here we consider 
misspecified time series models, and attempt to learn an ABC distance using Bayesian optimization
to improve forecasting performance for predictive distributions constructed from the ABC posterior.  
We target good forecasting performance out-of-sample, with this being assessed 
using a scoring rule appropriate for the problem.  
In addition to learning the distance in standard ABC, we also consider learning
a distance to improve forecasting performance for neural posterior estimation
with prior-fitted networks (NPE-PFN) \citep{vetter+ggm25}
with an ABC-like localization step.  Related preconditioned neural posterior
estimation approaches have been considered for misspecified simulators 
by \cite{kelly+fwd26}, but not in the context of time series forecasting.

In this work we consider methods based on summary statistics for the data, where 
the full data is projected into a lower-dimensional space before further analysis.  
The ABC distance is then defined in summary statistic space, and is often a weighted Euclidean distance, where 
the weights on different summaries need to be specified.    
We make two main contributions. First, we demonstrate that optimizing distance weights in ABC using a chosen scoring rule can improve the quality of forecasts
when evaluated in terms of the same scoring rule. Learning the weights enables downweighting irrelevant summaries and reducing the influence of summaries that are difficult to match due to misspecification.  
We further establish a connection between linear pooling for forecast combination and ABC with randomized distances, showing that empirical pooling weight estimation is a form of adaptive distance learning.
Second, we extend these ABC methods to SBI methods based on tabular foundation models and prior-data fitted networks \citep{vetter+ggm25}. Here, an ABC-type step provides a context set for in-context learning, and we demonstrate that learning distance weights in these algorithms similarly improves forecasting performance. 

There is substantial previous work on distance learning in ABC.  To the
best of our knowledge,
none of the existing work is in the context of forecasting misspecified time series.
Simple methods for summary statistic scaling for ABC can be thought of 
as implementing distance learning, for which the 
summary statistic weights are often based on 
features of the summary statistic prior predictive distribution.  These
scaling methods have been enhanced in many directions. \cite{jung+m11} 
employ a genetic algorithm to estimate summary statistic weights using a fitness function that incorporates ABC point estimation quality. \cite{prangle17} explored adjusting summary statistic scaling within iterative ABC algorithms such as population Monte Carlo.
\cite{gutmann+dkc18} develop a learned ABC discrepancy based on classification accuracy, while \cite{harrison+b20} propose selecting summary statistic weights to maximize posterior information gain. 
\cite{schalte+ah21} modify Prangle's approach to incorporate outlier-robust distances, and \cite{schalte+h23} simultaneously learn regression-based summary statistics and distances, considering weights on data points informed by the regression-based summaries.
\cite{thomas+slkcp25} describe an innovative generalized Bayesian approach that addresses model misspecification issues. Building on earlier Bayesian optimization-based LFI methods inspired by ABC 
\citep{gutmann+c16}, they consider additive discrepancies with terms for summary statistic blocks, 
scaling terms by minimum observed discrepancy values.

Since we consider learning the distance to improve forecasting for
misspecified time series, our work is also related to a large and
recently active literature
on LFI methods under misspecification.  
A common approach to mitigating the effects of misspecification 
in SBI considers a model expansion
using an ``error model'' that may involve additional parameters.  In ABC,
the kernel can be interpreted as a model error term \citep{wilkinson13} 
and it is commonly defined through a distance and one-dimensional kernel.  
From this perspective, an error model
with adjustable parameters is closely related to the motivation for an adaptive
ABC distance.  However, the specification of an error model in ABC 
is not straightforward \citep{schmon+ck20}.    There is much work on the use of error models for handling 
misspecification in other likelihood-free inference approaches, such as
Bayesian synthetic likelihood (BSL) \citep{frazier+d21}, and neural 
methods for likelihood and posterior estimation \citep{ward2022robust,kelly+nfwd24}. 
The error terms are designed to absorb misspecification, by allowing a sparse subset of summary
statistics to be ignored if they cannot be matched.
So far we have discussed likelihood-free methods using summary statistics, but there is an increasing
body of work on full-data distance methods which do not use summaries.  
These methods are often based on distributional divergences, such as maximum mean
discrepancy \citep{park+js16}, Kullback-Leibler divergence
\citep{jiang18}, Wasserstein distance \citep{bernton19}, energy distance
\citep{nguyen+alf20} and Cramer von Mises distance \citep{frazier20}.  Theoretical aspects
of full-data distance methods are considered in \cite{legramanti+da25}, and 
a recent review of the area 
is \cite{drovandi+f22}.  In some of their examples, \cite{drovandi+f22} use additive combinations of different
discrepancies with weights chosen adaptively based on robust measures of variability.    
 
None of the existing work on distance learning considers 
the explicit goal of improving forecasting performance in misspecified time series models, to the best of our knowledge.
Our work is inspired by recent works on the use of ABC methods 
for forecasting, such as \cite{frazier+mmm19} and \cite{weerasinghe+lmf23}.  
\cite{frazier+mmm19} consider ABC methods for parameter estimation, with summary statistic
choice guided by out-of-sample predictive performance for a scoring rule, and demonstrate
empirically and theoretically 
that even if the ABC posterior approximation is poor, predictive performance 
may be relatively insensitive to this if the ABC posterior gives consistent point
estimation and the sample size is large. They also consider state space models and observe that forecasting only requires filtering, not smoothing, after obtaining parameter samples using ABC.  \cite{frazier+mmm19} do not
consider the case of misspecified models, and this is addressed in 
\cite{weerasinghe+lmf23}.  They suggest choosing 
summary statistics using a flexible auxiliary model with closed form predictive distributions, 
and using the gradient of a scoring rule measure of predictive
performance at an optimal auxiliary model parameter value for the 
observed data to form the summaries.  
They also consider making
forecasts directly with the auxiliary model, using a generalized Bayes perspective where a loss likelihood
is constructed using the chosen scoring rule \citep{loaiza-maya+mf21}.  Variational approximations to the 
generalized Bayes posterior are also possible \citep{frazier+lmk25}.  
These existing works do not attempt to improve forecasting performance by distance learning, 
which is the focus of the current work.  There is much other
work on ABC methods for time series (e.g. \citealp{jasra+smm12}, \citealp{canale+r16},
\citealp{martin+mfmr19}, \citealp{mckinley+cd09}, \citealp{jarvenpaa+c23}) but here we focus on work explicitly 
concerned with model misspecification.  

In the next section we give an introduction to ABC and tabular foundational models with prior-data fitted
networks for SBI, and describe how the choice of a distance enters into these algorithms.  Section 3 
describes the approximate Bayesian forecasting methods of  \cite{frazier+mmm19} and 
\cite{weerasinghe+lmf23}, and how these are used in time series forecasting problems under misspecification.  
Section 4 describes our adaptive distance learning approach, and 
frames linear pooling for ABC forecast combination as adaptive learning of a
randomized ABC distance. Section 6 
considers several examples
and Section 7 gives concluding discussion.  

\section{ABC and NPE-PFN}\label{sec:LFI}

Before we discuss approximate Bayesian forecasting and 
adaptive distance learning, we introduce briefly the two LFI methods
used in this work, ABC \citep{sisson+fb18} 
and the NPE-PFN method of \cite{vetter+ggm25}.  

\subsection{Approximate Bayesian computation}

Suppose that there is a model with parameter $\theta$, data to be observed $y$
with a density $p(y|\theta)$, and observed data $y_{\text{obs}}$.  
We consider Bayesian inference with a prior density $\pi(\theta)$ for $\theta$.  
Let $S=S(y)$ be a mapping of $y$ into a lower-dimensional
space of summary statistic values.  The summary
statistic is chosen to be informative about $\theta$, and ABC methods
approximate the posterior density of $\theta$ given the observed
summary statistic value, which is denoted 
$S_{\text{obs}}=S(y_{\text{obs}})$.  

A widely-used rejection ABC method is given in Algorithm \ref{rejectionABC}.
It repeatedly generates synthetic summary statistic values from the Bayesian model, 
until one of these is within a tolerance distance 
$h$ of the observed summary, upon which
the corresponding parameter value is returned as an approximate posterior draw.  
A variant on this, given in Algorithm \ref{rejectionABCbatch}, draws $N$ parameter and summary statistic pairs from
the Bayesian model in a batch, and then chooses a quantile of distances from
the synthetic summaries to the observed value as the tolerance, to ensure a certain acceptance
rate in the algorithm.   

\begin{algorithm}
\caption{Rejection ABC algorithm}
\label{rejectionABC}
\begin{algorithmic}[1]
    \Statex \textbf{Inputs:} Prior density $\pi(\theta)$, tolerance $h>0$, summary statistic mapping $S=S(y)$, 
    observed summary statistic value $S_{\text{obs}}=S(y_{\text{obs}})$, 
    distance function $d(\cdot,\cdot)$ defined in the summary statistic space.
    \Statex \textbf{Output:} A sample from the ABC posterior distribution with tolerance $h$.
    \Repeat
      \State Simulate $\widetilde{\theta}\sim \pi(\theta)$.
      \State Simulate $\widetilde{S}\sim p(s|\widetilde{\theta})$, where $p(s|\theta)$ is the density of $S$ given $\theta$
    \Until{$d(\widetilde{S},S_{\text{obs}})<h$.}
    \State Return $\widetilde{\theta}$
\end{algorithmic}
\end{algorithm} 

\begin{algorithm}
\caption{Rejection ABC batch algorithm}
\label{rejectionABCbatch}
\begin{algorithmic}[1]
    \Statex \textbf{Inputs:} Prior density $\pi(\theta)$, number of samples $N$ drawn
    from the prior, fraction of accepted samples $\delta$, summary statistic mapping $S=S(y)$, 
    distance function $d(\cdot,\cdot)$ defined in the summary statistic space.
    \Statex \textbf{Output:}  A set of ABC posterior samples.
    \For{$i = 1$ \textbf{to} $N$}
      \State Simulate $\widetilde{\theta}_i\sim \pi(\theta)$.
      \State Simulate 
    $\widetilde{S}_i\sim p(s|\widetilde{\theta}_i)$, where $p(s|\theta)$
    is the density of $S$ given $\theta$.  
      \State Compute $d_i=d(\widetilde{S}_i,S_{\text{obs}}))$
    \EndFor
    \State Return $\{\widetilde{\theta}_i: 1\leq i\leq N, d_i\leq d^\delta\}$, where 
    $d^\delta$ is the lower $\delta$-quantile of $(d_1,\dots, d_N)$.
\end{algorithmic}
\end{algorithm} 

The choice of distance in these algorithms is often a Euclidean distance after some preliminary scaling
of the parameters, or a Mahalanobis distance using a summary statistic covariance matrix.  
This basic rejection ABC 
algorithm is rather inefficient and can be generalized in many ways, 
see \cite{fan+s18} for a summary and \cite{picchini+t23} for
recent developments.  
Methods like regression adjustment \citep{beaumont+zb02} have also 
been suggested to improve standard ABC methods, but simple ABC methods 
often behave quite well compared to more sophisticated variants when the model
is misspecified \citep{frazier+rr20}.  We will
also consider adapting the distance for a localized version of the NPE-PFN method of
\cite{vetter+ggm25}, which we explain next.

\subsection{Neural posterior estimation with prior-data fitted networks}

\cite{vetter+ggm25} consider neural posterior density estimation with prior-data 
fitted networks (NPE-PFN) for performing SBI, which is implemented
using the tabular foundation model TabPFNv2 \citep{hollman+mpkkhsh25}.  Hereafter
references to TabPFN mean TabPFNv2.  
TabPFN allows estimation of a predictive distribution in problems of the following kind.  
Suppose there is a tabular dataset consisting of response and feature vector pairs
${\cal D}=\{(y_i,x_i); i=1,\dots, n\}$, where the responses can be real-valued (in regression problems)
or class labels (for classification problems).   We are interested in generating a predictive density for
a target feature vector $x_0$ (or a set of such target feature vectors) given the training data.  TabPFN approximates the predictive
distribution for the response at the test feature vectors directly.  We pass the pair
$({\cal D},x_0)$ (the context) 
to TabPFN, which returns a predictive distribution for 
$y_0$.  TabPFN is an acronym for ``tabular prior-data fitted network'' and here prior-data fitted means
that no training is done involving the context data.  
Similar in-context learning (ICL) methods are used widely 
in modern machine learning, with large language models (LLMs)
a common example.  TabPFN can be thought of as doing  
approximate Bayesian inference based on a very flexible class of models.  
It is pre-trained using a large number of synthetic datasets (over 100 million).  
The synthetic data are generated using 
structural causal models (see, for example, \citealp[Section 3.1]{peters+js17}), which are specified through a graph and possibly 
nonlinear functions defined at the nodes
specifying dependence on parents in the graph and noise.   
The synthetic datasets have different numbers
of data points and features.  A transformer model \citep{vaswani+spujgkp17}
is used to define a mapping of a particular dataset and set of test feature vectors to predictive
distributions for the label at the test features, with a cross-validatory criterion using a logarithmic
scoring rule optimized in training.  The standard approach also considers
transformations of the features as well as 
various post-processing adjustments, and has the capacity to deal with 
missing features and outliers in the features and labels.
For a good high-level overview of the basic ideas for statisticians see \cite{zhang+ttl25};  
these authors also document the impressive performance of the approach as an off-the-shelf
predictive tool in several applications.  
Further technical details can be found in \cite{hollman+mpkkhsh25}, where it is claimed that
the method ``yields dominant performance for datasets with up to 10,000 samples and 500 features''.  
A recent development is the release of TabPFN-2.5 \citep{grinsztajn25}, the successor of TabPFNv2, 
where it is claimed that the superior performance of TabPFNv2 compared to other benchmarks 
extends to datasets with up to 50,000 
samples and 2,000 features.  In this work we focus on adaptive distance learning for the neural
posterior estimation (NPE) approach of \cite{vetter+ggm25}, which is implemented using TabPFNv2.  

The NPE-PFN approach of \cite{vetter+ggm25} starts with a training set of prior parameter and data set pairs.  Here we consider the use of summary statistics, so we have prior parameter and summary statistic
pairs.  So 
\begin{align}
  {\cal D} & =\{(\theta_i,S_i)\sim \pi(\theta)p(s|\theta); i=1,\dots, n\}. \label{fullD}
\end{align}
We want to estimate
from this dataset the predictive density $p(\theta|S_{\text{obs}})$ for $\theta$ given the observed
summary statistic $S_{\text{obs}}$.  However, TabPFN cannot do this directly, since 
it performs only univariate prediction.  To address this issue \cite{vetter+ggm25} consider 
the components of the parameter vector in order.  They decompose the joint posterior as
\begin{align}
p(\theta|S_{\text{obs}}) & =\prod_{i=1}^p p(\theta_i|\theta_{<i},S_{\text{obs}}),  \label{decomposition}
\end{align}
where we have written $p$ for the dimension of $\theta$ and $\theta_{<i}$ for the vector of 
components of $\theta$ with index less than $i$, with $\theta_{<1}$ defined as a null vector.  
Now we can approximate the univariate densities in the product on the right-hand side
of \eqref{decomposition}.  To do so here, the method of \cite{vetter+ggm25} would
first consider a training dataset 
${\cal D}_1=\{(\theta_{i1},S_i); i=1,\dots, n\}$, where we consider only the first component of the
vector $\theta$
from \eqref{fullD} as the response.  From this, TabPFN can produce an estimate of the marginal
posterior $p(\theta_1|S_{\text{obs}})$, say $q(\theta_1|S_{\text{obs}})$.  Next, consider the dataset 
${\cal D}_2=\{(\theta_{i2},(\theta_{i1},S_i)); i=1,\dots,n\}$, considering $\theta_2$
as the response to be predicted, and $\theta_1$ as part of the feature vector together with the
summary statistic.  Then we can obtain, using TabPFN, an estimate of the conditional posterior, 
$p(\theta_2|\theta_1,S_{\text{obs}})$, say $q(\theta_2|\theta_1,S_{\text{obs}})$.  Continuing in this way, we approximate $p(\theta|S_{\text{obs}})$ as
$$p(\theta|S_{\text{obs}})=\prod_{i=1}^p p(\theta_i|\theta_{<i},S_{\text{obs}})\approx \prod_{i=1}^p q(\theta_i|\theta_{<i},S_{\text{obs}}).$$

\cite{vetter+ggm25} outline
a number of refinements.  The NPE-PFN has some strong advantages over alternative methods when
the training data set of simulations is small, which tends to be the case for simulators which are
computationally expensive.  However, the recommended limit of training sets for TabPFNv2 to
sample size less than $10,000$ means that for larger simulation budgets, NPE-PFN may not perform well.  
As a solution to this problem, one can include an ABC-type rejection step to include in the context
set only points $\{(\theta_i,S_i); d(S_i,S_{\text{obs}})<\epsilon\}$, where $\epsilon$ is chosen so
that the context set is of size less than $10,000$.  Later we will consider distance learning
with in-context learning of this type, referred to as localization or retrieval in the machine learning
literature (see, for example, \citet{nagler23}).  
\cite{vetter+ggm25} also consider a so-called truncated sequential version of 
their method, TS-NPE-PFN, where instead of drawing samples from the prior in the training set, an attempt
is made to truncate the prior to the region of the posterior support.  This means that less simulation
is required because we only simulate in the region of parameter space where there is a good chance
of getting simulated datsets consistent with the observed data.  
We do not consider the sequential approach here, since it is not obvious how to combine this with
distance learning under misspecification, but this is an interesting future research direction.  

\section{Approximate Bayesian forecasting}

Our work builds on the approximate Bayesian forecasting (ABF) framework
of \cite{frazier+mmm19} and \cite{weerasinghe+lmf23}, and we explain
this next. 
Let $Y=\{Y_t; t=1,2,\dots, \}$ be a time series.  We write
$Y_{\leq n}=(Y_1,\dots, Y_n)^\top$ for the first $n$ observations
of $Y$.  There is a model for $Y_{\leq n}$ with parameters
$\theta$ and density $p(y_{\leq n}|\theta)$.  The posterior density
given $y_{\leq n}$ is 
$$\pi(\theta|y_{\leq n})\propto \pi(\theta)p(y_{\leq n}|\theta),$$
where $\pi(\theta)$ is the prior density.
The predictive density for $Y_{n+1}$ given $Y_{\leq n}=y_{\leq n}$ is 
\begin{align}
  p(y_{n+1}|y_{\leq n}) & = \int p(y_{n+1}|y_{\leq n},\theta)
\pi(\theta|y_{\leq n})\,d\theta.  \label{one-step-pred}
\end{align}
\cite{frazier+mmm19} and \cite{weerasinghe+lmf23}, 
suggest to use an ABC 
approximation to the posterior density instead of the true
posterior density in \eqref{one-step-pred}.    
\cite{frazier+mmm19} explore the phenomenon of merging of
predictive distributions, which means here that
even if the ABC posterior is not a good approximation to the 
exact posterior, predictive distributions can still be similar 
provided there is posterior consistency for both the ABC and exact 
posterior densities and if $n$ is sufficiently large.
Summary statistics may be chosen based on a measure of predictive loss. 
See also \cite{lacoste-julian+hg11}, \cite{loaiza-maya+mf21} and \cite{frazier+lmk25} for other recent work on
loss-based Bayesian prediction.  
We will write $\pi_h(\theta|y_{\leq n})$ for the ABC posterior with distance tolerance $h$.  Also write  
\begin{align}
  p_h(y_{n+1}|y_{\leq n}) & = \int p(y_{n+1}|y_{\leq n},\theta)
  \pi_h(\theta|y_{\leq n})\,d\theta, \label{abc-one-step-pred}
\end{align}
for the predictive density \eqref{one-step-pred} with the 
posterior density replaced by the ABC posterior.

State space models (SSMs) are a particularly interesting class of models 
for approximate Bayesian forecasting.  
\cite{frazier+mmm19} considered SSMs in their work, but only for
the case of correctly specified models.  More recently, 
\cite{weerasinghe+lmf23} extend 
\cite{frazier+mmm19} by discussing the implementation
of approximate Bayesian forecasting when the assumed SSM
is misspecified.  We focus on the misspecified case
in our work also.  
  
A state space model with observations $Y_t$ and latent 
states $Z_t$ for $t=1,2,\dots$, can be specified as
\begin{align}
  Y_t|Z_t,\theta  & \sim f_t(y_t|z_t,\theta),  \label{obs} \\
  Z_t|Z_{t-1},\theta & \sim g_t(z_t|z_{t-1},\theta),  \label{state}
\end{align}
for $t=1,2,\dots$, 
where the initial value $Z_0$ has prior density $\pi(z_0|\theta)$.  
The observations are conditionally independent given the states, 
and the states evolve according to a first order Markov process. 
The posterior density for $\theta$ is difficult to compute in this model, 
since the states need to be integrated out to obtain the likelihood
$p(y_{\leq n}|\theta)$.  Exact posterior sampling
for nonlinear and non-Gaussian SSMs usually involves the use of particle MCMC
\citep{andrieu+dh10} or pseudo-marginal Metropolis-Hastings
algorithms \citep{beaumont03,andrieu+r09}, 
which are computationally intensive and 
infeasible with a high-dimensional state vector.
Replacing the exact posterior density
$\pi(\theta|y_{\leq n})$ with an ABC approximation $\pi_h(\theta|y_{\leq n})$
may be particularly attractive in the case of SSMs, since likelihood
computations are not required for ABC.    
In SSMs, we can write 
\begin{align}
  p(y_{n+1}|y_{\leq n},\theta) & =\int f_{n+1}(y_{n+1}|z_{n+1},\theta)
g_{n+1}(z_{n+1}|z_n,\theta)p(z_n|y_{\leq n},\theta)dz_{\leq n+1},  \label{ssmlike}
\end{align}
where $f_{n+1}(y_{n+1}|z_{n+1},\theta)$ comes from \eqref{obs}, $g_{n+1}(z_{n+1}|z_n,\theta)$ from \eqref{state}, 
and $p(z_n|y_{\leq n},\theta)$ is the density of $Z_n$ given $y_{\leq n},\theta$.  
It is possible to draw samples from $p(y_{n+1}|y_{\leq n},\theta)$ using a particle
filter \citep{gordon+ss93}. 
Substituting \eqref{ssmlike} into \eqref{one-step-pred} or 
\eqref{abc-one-step-pred}
allows approximation of one-step-ahead predictive densities, by
drawing samples $\theta$ from the posterior or its approximation, 
drawing $z_n$ from the density of $Z_n|y_{\leq n},\theta$, and then
averaging the first two terms of the product in the integrand 
at \eqref{ssmlike} over
the draws.  Samples can also be drawn from $f(y_{n+1}|z_{n+1},\theta)$
or $g_{n+1}(z_{n+1}\mid z_n,\theta)$ if 
these terms are intractable.  \cite{frazier+mmm19} observe that
constructing predictive density approximations requires only filtering
to draw samples for $Z_n$ given $y_{\leq n},\theta$, and not smoothing. 

\subsection*{Summary statistic choice for ABF and misspecified SSMs}

A central idea of the loss-based approximate Bayesian forecasting approach
of \cite{weerasinghe+lmf23} is to choose
summary statistics with a particular measure of forecasting accuracy
in mind.  The measure of forecasting accuracy will be defined through
a scoring rule, and we explain this first.  
Let $\mathcal{P}$ be a set of distributions, and
$\mathcal{Y}$ a set of possible values for an observation.  
A scoring rule is
a function $Q:\mathcal{P}\times \mathcal{Y}\rightarrow \mathbb{R}\cup \{-\infty,\infty\}$, 
where $Q(F,y)$ is the reward for forecasting the observation 
$y\in\mathcal{Y}$ by the distribution $F\in \mathcal{P}$.
We follow the convention that a larger value is desirable (a 
positively-oriented scoring rule).  \cite{gneiting+r07} give an extensive
summary of the literature on scoring rules.

Consider a random observation $Y\sim P\in \mathcal{P}$, 
and write the expected score (which is assumed to be well-defined) by
$$Q(F,P)=E(Q(F,Y)).$$
A scoring rule is said to be proper if for every $P,F\in \mathcal{P}$,
$Q(P,P)\geq Q(F,P)$.  For a proper scoring rule the expected 
reward is maximized
by using $P$ as the forecast if the true distribution is $P$.  
A scoring rule is strictly proper with respect to $\mathcal{P}$ if
$Q(P,P)=Q(F,P)$ implies that $F=P$.  A strictly proper scoring rule
encourages honesty (i.e. to maximize the expected
reward the forecaster should
choose the forecast distribution according to their true beliefs).  
We will consider forecasting
for continuous quantities, where the forecast distribution $F\in \mathcal{P}$
has a density $f(y)$ say.  Following
\cite{weerasinghe+lmf23}, we use four different
scoring rules in the examples.  The first is the logarithmic
score \citep{good52},
\begin{align}
  Q_{\text{LS}}(F,y) & = \log f(y).  \label{logarithmic}
\end{align}
The second is the continuous ranked probability score (CRPS) \citep{brown74}, 
\begin{align}
  Q_{\text{CRPS}}(F,y) & = -\int (F(z)-I(z\geq y))^2\,dz.
\end{align}
The third is the interval score \citep{dunsmore68}
\begin{align}
 Q_{\text{IS}}^\rho(F,y) & = (u_\rho-l_\rho)+\frac{2}{\rho}(l_\rho-y)I(y<l_\rho)+\frac{2}{\rho}(y-u_\rho)I(y>u_\rho),
\end{align}
where interest focuses on a $(1-\rho)\times 100\%$ prediction interval for the
forecast distribution $F$ and $l_\rho$ and $u_\rho$ are the $\rho/2$ and $1-\rho/2$ quantiles of $F$. 
Finally, the censored log score \citep{diks+pv11} is
\begin{align}
  Q_{\text{CLS}}(F,y) & = \log f(y)I(y\in C)+\log \int_{z\notin C} f(z)\,dz I(y\notin C).
\end{align}

We now discuss the approach of \cite{weerasinghe+lmf23} for choosing
summary statistics in ABF based on a measure of predictive
performance in the form of a scoring rule.  
Their approach is particularly valuable when the assumed model is 
misspecified.  They construct summary statistics
by making use of an auxiliary 
model \citep{gleim+p13,drovandi+pl15,martin+mfmr19}, 
but in a way that makes
explicit use of the forecasting performance of the auxiliary model.
  
Write $p_A(y_n|y_{\leq n-1},\eta)$ for the one-step ahead
predictive density of $Y_n$ given $y_{\leq n-1},\eta$ for
any integer $n\geq 1$ for a chosen auxiliary
model with parameters $\eta$.   The auxiliary model
is chosen so that a closed-form expression for 
this predictive density exists.  We write $P^{(n)}_A(\eta)$ for
the corresponding distribution.  The closed form for the predictive
densities is important for computing scoring rules and their
gradients which is used in the summary statistic construction
below.  This is why an auxiliary model is used and not the assumed
SSM directly. 
    
For a set of training observations of length $T$, $y_{\leq T}$, 
the one-step ahead predictive performance can be summarized 
by 
\begin{align}
  q(\eta,y_{\leq T}) & =n^{-1} \sum_{n=1}^T Q(P^{(n)}_A(\eta),y_n),  \label{Qtrain}
\end{align}  
using a scoring rule $Q$ chosen to reflect the predictive goals
of the analysis.
Now define
$\widehat{\eta}(y_{\leq T})=\arg \max_\eta q(\eta,y_{\leq T})$
for the value of $\eta$ maximizing the measure \eqref{Qtrain} of predictive
performance.
Suppose that $w_{\leq T}$ is a simulated time series from the
assumed statistical model (not the auxiliary model).  
Then \cite{weerasinghe+lmf23} define summary statistics for $w_{\leq T}$
based on the auxiliary model by
\begin{align}
  S(w_{\leq T}) & = \left. \frac{\partial}{\partial \eta} q(\eta,w_{\leq T})\right|_{\eta=\widehat{\eta}(y_{\leq T})}. \label{auxiliaryS}
\end{align}
Note that the gradient is computed at $\eta=\widehat{\eta}(y_{\leq T})$ even 
when considering the data $w_{\leq T}$ so that the optimization to
find $\widehat{\eta}(y_{\leq T})$ only needs to be performed once.
For the observed data $y_{\leq T}$, we have $S(y_{\leq T})=0$.  

The auxiliary model summary statistics capture features of
the data which are important for good forecasting performance
in terms of the scoring rule $Q$.  \cite{weerasinghe+lmf23} compare
ABF with the summary statistics \eqref{auxiliaryS} with forecasting
based on the misspecified auxiliary model directly, as well as a generalized 
Bayesian approach employing the scoring rule
loss \eqref{Qtrain}.  They conclude that the ABF approach is
highly competitive, superior to using the auxiliary model
directly, and often better than the generalized
Bayesian loss-based forecasting method of \cite{loaiza-maya+mf21}.  
ABF also exhibits coherent prediction, in the sense that the best forecasts
result for a given scoring rule when the summary statistics were constructed
using the same scoring rule.

\section{Adaptive distance learning}

The approach of \cite{weerasinghe+lmf23} attempts to mitigate misspecification
by considering a scoring rule relevant to the problem at hand, and then choosing
summary statistics that are adapted to that rule.  Here we instead suggest to learn
an adaptive distance in ABC (or in NPE-PFN with localization) where distance weights
are chosen to optimize forecasting performance.  

In a correctly specified
model, it is optimal to implement ABC algorithms to approximate
the true posterior as closely as possible given the available computational
resources.  However, under misspecification, improved forecasting performance may result from choosing the weights in an ABC distance to
discard information.  
ABC distance learning enables discarding irrelevant summaries, as well as discarding
summaries that are hard to match.   

We parametrize
the ABC distance used in 
terms of weights $\omega=(\omega_1,\dots, \omega_J)^\top$, 
where $J$ is the summary statistic dimension.  Write $\Omega$ for the diagonal matrix with
diagonal elements $\omega$.  
Let $\Sigma$ be the covariance matrix of summary statistics $S$ drawn from the prior.  
For summary statistics $S$ and $S'$, we consider
a Mahalanobis distance, 
\begin{align}
 d_\omega(S,S') & =\left\{(S-S')^\top \Omega\Sigma^{-1}\Omega (S-S')\right\}^{1/2}. \label{mdist}
\end{align}
We fix one of the weights to one, without loss of generality say $\omega_1=1$, 
to fix the overall scale of the distance. This ensures 
changes to the ABC tolerance $h$
are not equivalent to a change of all distance weights by a common
multiplicative factor.
A small weight $\omega_j$ decreases the relative importance
of the $j$th summary statistic, with
$\omega_j=0$ removing the summary entirely.  Although adapting an ABC 
distance function is not new, the novelty of our approach lies in 
choosing the distance weights to optimize forecasting performance in 
approximate Bayesian forecasting.  

It is assumed in what follows that in addition to the training
set observations $y_{\leq T}$, we have a validation set
$\widetilde{y}_j=y_{T+n}$, $n=1,\dots, \widetilde{T}$, and observations
for a forecasting period $\breve{y}_n=y_{n+T+\widetilde{T}}$, 
$n=1,\dots, \breve{T}$.  
Write $\widetilde{y}_{\leq \widetilde{T}}=(\widetilde{y}_1,\dots, \widetilde{y}_{\widetilde{T}})^\top$, and $\breve{y}_{\leq \breve{T}}=(\breve{y}_1,\dots, \breve{y}_{\breve{T}})^\top$.  We use $y_{\leq T}$ for estimating the ABC posterior, 
$\widetilde{y}_{\leq \widetilde{T}}$ for tuning the ABC distance weights, 
and $\breve{y}_{\leq \breve{T}}$ for evaluating the forecasting performance.

For any given value of the weights $\omega$, we can estimate an ABC
posterior using the distance $d_\omega(\cdot,\cdot)$ and the observations
$y_{\leq T}$.  Suppose we have ABC posterior samples
$\theta^{(\omega,i)}$, $i=1,\dots, I$.  
Write $P^{(n)}(\theta)$ for the SSM predictive distribution for 
$Y_n$ given $y_{\leq n-1},\theta$, with density
$p(y_n|y_{\leq n-1},\theta)$.  We measure the predictive performance
on the validation set by  
\begin{align}
  \widetilde{q}(\omega) & = \widetilde{T}^{-1}\sum_{n=1}^{\widetilde{T}}Q\left(I^{-1}\sum_{i=1}^I  P^{(T+n)}(\theta^{(\omega,i)}),\widetilde{y}_n\right),  \label{validationQ}
\end{align}
where the scoring rule $Q$ is the same one used in the definition
of the summary statistics, if the method of \cite{weerasinghe+lmf23}
is used to construct the summaries.  In practice, $P^{(n)}(\theta)$ or its density
need to be approximated numerically.  
Our goal is to optimize \eqref{validationQ} to obtain the final
distance weights $\omega^*=\arg \max_\omega \widetilde{q}(\omega)$.  These weights are then
used to perform forecasting for the observations $\breve{y}_{\leq \breve{T}}$
in the forecast period.  With the Monte Carlo approximation of the
ABC posterior in \eqref{validationQ} and possibly Monte Carlo approximation of the 
predictive distributions in the case of SSMs, we only have noisy evaluations of a performance
measure to use for optimization.  
To address this issue, we use Bayesian optimization (BO) 
\citep{garnett23}
to optimize the weights. BO methods can deal 
with optimizing computationally expensive objectives where only noisy
function evaluations without gradients are available.
In this work we use the implementation of Bayesian optimization
given in the BoTorch package \citep{balandat20}. 

If the posterior distribution is not produced by ABC but instead using the NPE-PFN method, 
the basic method is similar, but now we have samples 
$\theta^{(\omega,i)}$, $i=1,\dots, I$ produced using the NPE-PFN method.  Starting with
an initial set of samples $(\theta_i,S_i)$, $i=1,\dots, T$ from the joint
Bayesian model write $d_i^\omega=d_\omega(S_i,S_{\text{obs}})$, and $d^\omega_{(i)}$ for the order statistics. 
Then for the context set for the NPE-PFN method, we use $\{(\theta_i,S_i): d^\omega_i\leq d^\omega_{(1,000)}\}$, 
i.e. we choose the 1,000 initial samples with simulated summaries closest to the observed summary statistic, 
in terms of the distance measure.  Then with this context set we estimate the posterior, and generate
$\theta^{(\omega,i)}$, $i=1,\dots, I$, from that posterior density.  For the posterior for a given $\omega$, 
we can measure predictive performance on the validation set similar to \eqref{validationQ} but applied
to the data in the forecasting period $\breve{y}_{\leq \breve{T}}$, and 
we use Bayesian optimization to find the optimal weights based
on $\widetilde{y}_{\leq \widetilde{T}}$.   

\subsection{Linear pooling as adaptive learning of a randomized distance}

We now describe an interesting connection between adaptive learning of a (randomized)
ABC distance, and combining predictive densities by linear opinion pooling \citep{stone1961}, so that learning weights in the opinion pool 
can be thought of as adaptive distance learning.  
Suppose that we have $K$ different possible choices of the summary
statistics, $S^{(j)}$, $j=1,\dots, K$.  The corresponding observed values are $S_{\text{obs}}^{(j)}=S^{(j)}(y_{\text{obs}})$.  We use a distance $d^{(j)}(\cdot,\cdot)$ for the summary statistic
$S^{(j)}$, and this distance is not learnt adaptively.  For example, 
the distance might take the form of the Mahalanobis distance in 
\eqref{mdist} with $\Omega$ fixed at the identity matrix.  
For the $j$th summary statistic vector we can construct a $1$-step ahead predictive distribution 
$p^{(j)}(y_{n+1}\mid y_{\leq n})$:
$$p^{(j)}(y_{n+1}\mid y_{\leq n})=
\int p(y_{n+1}|y_{\leq n},\theta)\pi_{h_j}(\theta|S_{obs}^{(j)})\,d\theta,$$
where $\pi_{h_j}(\theta|S_{obs}^{(j)})$ is the ABC posterior using summary statistic $S^{(j)}$, 
distance $d^{(j)}(\cdot,\cdot)$ and tolerance $h_j>0$.  
A linear opinion pool of the predictive densities for different summaries with non-negative weights $\omega=(\omega_1,\dots, \omega_K)^\top$, $\sum_{j=1}^K \omega_j=1$, is
\begin{align*}
 p^{1:K}_{h,\omega}(y_{n+1}\mid y_{\leq n})& =\sum_{j=1}^K \omega_j p^{(j)}(y_{n+1}\mid y_{\leq n}) \\
  & = \sum_{j=1}^K \omega_j \int p(y_{n+1}|y_{\leq n},\theta)\pi_{h_j}(\theta|S_{\text{obs}}^{(j)})\,d\theta \\
  & = \int p(y_{n+1}|y_{\leq n},\theta) \sum_{j=1}^K \omega_j \pi_{h_j}(\theta|S_{\text{obs}}^{(j)})\,d\theta,
\end{align*}
which is the predictive density integrating out $\theta$ using the pooled posterior density
\begin{align}
 \pi^{1:K}_{h,\omega}(\theta\mid y_{\leq n}) & := \sum_{j=1}^K \omega_j \pi_{h_j}(\theta\mid S_{\text{obs}}^{(j)}). \label{abcpooled}   
\end{align}
Pooled predictive densities like this have been considered in \cite{frazier+dkn25} for ABC with different
summary statistics and by \cite{yao+bd24} for SBI.  
Drawing samples from the pooled posterior density \eqref{abcpooled}
can be thought of as implementing an ABC algorithm, but where steps 2-4 in Algorithm \ref{rejectionABC}
are replaced by a distance comparison involving a random choice of the summary statistic
distance.  The new algorithm is given in Algorithm \ref{randomizedABC} below, and it is straightforward
to verify that this produces a sample from \eqref{abcpooled}.  
\begin{algorithm}
\caption{ABC pooled posterior sampling algorithm}
\label{randomizedABC}
\begin{algorithmic}[1]
    \Statex \textbf{Inputs:} prior density $\pi(\theta)$, summary statistic mappings $S^{(j)}=S^{(j)}(y)$ and 
    observed summary statistic values $S^{(j)}_{\text{obs}}=S^{(j)}(y_{\text{obs}})$, distance function $d^{(j)}(\cdot,\cdot)$
    defined in the space of the summary statistic $S^{(j)}$, $j=1,\dots, K$.  Vector of non-negative weights $\omega=(\omega_1,\dots, \omega_K)^\top$, $\sum_{j=1}^K \omega_j=1$.  
    \Statex \textbf{Output:} A sample from the pooled ABC posterior distribution \eqref{abcpooled}.
    \State Draw $j=l$ with probability $\omega_l$, $l=1,\dots, K$.
    \Repeat
      \State Simulate $\widetilde{\theta}\sim \pi(\theta)$.
      \State Simulate $\widetilde{y}\sim p(y|\widetilde{\theta})$.  
      \State Compute $\widetilde{S}^{(j)}=S^{(j)}(\widetilde{y})$,
    \Until{$d^{(j)}(\widetilde{S}^{(j)},S^{(j)}_{\text{obs}})<h_j$.}
    \State Return $\widetilde{\theta}$.
\end{algorithmic}
\end{algorithm} 
Since linear opinion pooling of the predictive densities is just ordinary ABC prediction
with a randomized distance, learning the
pooling probabilities is a type of adaptive distance learning.  This is why we have overloaded 
the $\omega$ notation used previously for summary statistic weights.  The pooling probabilities can be
estimated in a similar fashion to our previous discussion, based on a validation set.  Interestingly, 
there is a precedent for considering a randomized ABC distance in Bayesian inference
for misspecified models:  
\cite{miller+d19} consider an ABC-like conditioning involving relative entropy and a randomized
neighbourhood size in a ``coarsened posterior'' for providing robustness of posterior
inferences to perturbations of an assumed Bayesian model.
Linear opinion pooling in the context of probabilistic forecasting is also related to 
Bayesian model stacking \citep{yao+vsg18}.  

There is a large literature on probabilistic forecast combination, and unsurprisingly 
not all such combination
methods can be thought of as adaptive distance learning in the ABC context.  
Since our focus in this article is about adaptive distance learning, we will focus on simple
linear pooling, but more sophisticated methods for probabilistic forecast combination exist.
One disadvantage of linear pools is that the variance of the combined forecast 
is larger than the corresponding weighted
sum of variances of the component densities, meaning that the pooled forecast is overdispersed
compared to the component forecasts in that sense, which may not always
be appropriate but can be good if component forecasts are overconfident.  
For a recent survey of forecast combination methods, including a discussion of
linear pooling and its variants for probabilistic forecasting, see \cite{wang+hfk23}. 
Although we attempt to learn optimal weights, it has been noted that using
equal weights is often competitive with more complex methods for forecast combination
with linear pools, both for point forecasting and probabilistic forecasts, 
particularly when data for estimating the weights is limited.  
This has been called the ``forecast combination puzzle'' \citep{stock+w04}.  It is possible to also make the linear pooling weights time-varying, but we do not consider this here.  

When learning pooling weights based on $\widetilde{y}_{\leq \widetilde{T}}$, 
we need to obtain a sample $\theta^{(\omega,i)}$, $i=1,\dots, I$, 
from the pooled posterior \eqref{abcpooled}.  
Algorithm \ref{randomizedABC} above requires specifying the tolerances
$h_j$, $j=1,\dots, K$.  It is easier to draw prior samples in a batch and
specify an acceptance rate, and in practice we sample from the pooled
posterior using a variant similar to Algorithm \ref{rejectionABCbatch}
adapted to the pooled case.  
We first draw an ABC posterior sample of size $M=\lfloor N\delta \rfloor$ 
using Algorithm \ref{rejectionABCbatch}
for each of the summary statistics $S^{(j)}$ with corresponding
distance $d^{(j)}(\cdot,\cdot)$, $j=1,\dots, K$.  Write the samples 
for summary statistic $j$ as $\theta^{(j)}_m$, $m=1,\dots, M$, $j=1,\dots, K$. 
Then for
$i=1,\dots, I$ we draw each pooled sample by:  1) Draw $j$ with probability $\omega_j$, $j=1,\dots, K$, 2) Draw $\theta^{(\omega,i)}$ uniformly at random from
$\theta^{(j)}_m$, $m=1,\dots, M$.  
We have not yet discussed how to obtain the different summary statistics
$S^{(j)}$, $j=1,\dots, K$.  In our later examples, we do this by making
use of auxiliary summary statistics for different scoring rules - we choose 
$K=3$ and the auxiliary summary statistics of \cite{weerasinghe+lmf23} 
constructed using the LS, CLS$_{20}$ and CLS$_{80}$ scoring rules.

\section{Examples}\label{sec:Example}

We now examine the empirical performance of our approaches for simulated
and real examples considered in \cite{weerasinghe+lmf23}.

\subsection{A stochastic volatility model}

Our first example considers a stochastic volatility model of the form
\begin{align}
  y_t & = \mu_y+\exp(z_t/2)\epsilon_y, \;\;\;\epsilon_y\sim N(0,1),  \label{obseq} \\
  z_t & = \mu_z+\phi(z_{t-1}-\mu_z)+\sigma_z \epsilon_z, \;\;\;\epsilon_z\sim N(0,1),  \label{state-eq}
\end{align}
where $z_1\sim N(\mu_z,\sigma_z^2/(1-\phi^2))$.  The unknown parameters are 
$\theta_a=(\mu_y,\mu_z,\phi,\sigma_z)^\top$.  Equations \eqref{obseq} and 
\eqref{state-eq} are the assumed model
for the analysis (the subscript in $\theta_a$ is for ``assumed''), but we consider observed data which is simulated from a true data generating
process (DGP) which differs from this.  Let $w_t$ be defined by
\begin{align}
  w_t & = \exp(z_t/2)\epsilon_y, \;\;\;\epsilon_y\sim N(0,1),  \label{obseq-dgp} \\
  z_t & = m_z+f_z(z_{t-1}-m_z)+s_z \epsilon_z, \;\;\;\epsilon_z\sim N(0,1),  \label{state-eq-dgp}
\end{align}
with $z_1\sim N(m_z,s_z^2/(1-f_z^2))$ and where $m_z$, $f_z$ and $s_z$ are unknown parameters.  
Then we consider $y_t$ that is a transformation of $w_t$, 
\begin{align}
  y_t & =G_{\zeta}^{-1}(F_w(w_t;m_z,s_z,f_z)), \label{transform}
\end{align}
where $F_w(\cdot;m_z,s_z,f_z)$ is the distribution function of the stationary distribution of $w_t$, and
$G_\zeta(\cdot)$ is the distribution function of a standardized skew-normal random variable
with shape parameter $\zeta$.  
The unknown parameters in the true DGP are denoted $\theta_d=(m_z,f_z,s_z,\zeta)^\top$.  
The model above for $w_t$ is of the same form as
that in \eqref{obseq} and \eqref{state-eq} with $\mu_y$ set to zero, and the transformation 
of $w_t$ to $y_t$ makes
the marginal distribution of $y_t$ standardized skew-normal with shape parameter $\zeta$, 
giving some control over the skewness, unlike the assumed model.
The distribution function $F_w(w_t;m_z,f_z,s_z)$ can be estimated by simulation.  

\subsubsection{Simulation details}

In the simulations below, following \cite{weerasinghe+lmf23}, we generate the
observations using $(m_z,f_z,s_z)=(-0.4581,0.9,0.4173)$ in \eqref{obseq-dgp} and 
\eqref{state-eq-dgp}.  \cite{weerasinghe+lmf23} 
considered different values of 
$\zeta$, corresponding to different levels of misspecification.  We consider only the most
misspecified case of $\zeta=-5$.  We will focus on comparing forecast performance 
for two different summary statistic choices, two different forecasts lead times, using an adaptive
distance or fixed distance, using ABC or TabPFN to obtain approximate posterior samples, and
using the different scoring rules discussed in \cite{weerasinghe+lmf23} and in Section 3.  When considering a certain scoring rule for evaluating predictive
performance, we only consider auxiliary model summary statistics constructed
using the same scoring rule, if summaries are chosing using the approach
of \cite{weerasinghe+lmf23}.  Given their previous work, 
we take it as given that this is the best
choice of auxiliary model summary statistic.

For ABC posterior estimation, we use Algorithm \ref{rejectionABCbatch}, with $N=100,000$ and choose 
$\delta$ so that $I=100$ in \eqref{validationQ} within 
the Bayesian optimization for learning the distance weights.   
A small value of $I$ is chosen because in 
the weight optimization we must evaluate many sets of weights, and
the particle filter is the most computationally expensive part of the forecasting, 
which needs to be done for every posterior sample.  
Bayesian optimization is well-suited to optimizing a noisy objective.
Similar settings are used to the above 
for $N$ and $I$ when TabPFN is used to generate posterior samples.  

In our simulated data, we use $T=5000$, $\widetilde{T}=1000$ and $\breve{T}=1000$.  In approximating
the predictive densities $P^{(n)}(\theta)$ in \eqref{validationQ}, we use a boostrap
particle filter \citep{gordon+ss93}, with $P=100$ particles during the Bayesian optimization evaluation of the predictive performance of 
the optimal weights in the
forecast period.  Similar to the choice of $I$, a small value of $P$ is used
to control the computational cost of the
particle filter calculations for learning weights, with the Bayesian
optimization approach being tolerant of the noise
this introduces.  In addition, when running the particle filter
we start at time $T-1000$ to sample the states 
within the validation and forecasting periods.  Similar 
``windowing'' approaches to estimating the states
are used in subsampling methods for Bayesian inference in state space models \citep{aicher+pnff25}.  

Two choices for the summary statistics $S$ are considered:  1) auxiliary-model based summaries
as described in Section 3, defined according to the scoring rule used in the forecast evaluation, 
with the ARCH(1) auxiliary model discussed in \cite{weerasinghe+lmf23}, and leading
to a three-dimensional summary statistic;  and 2) 
a na\"{i}ve choice of $S$ of dimension 6, consisting of sample 
autocovariance values for $\{Y_t;t=0,1,\dots, T\}$ and
$\{Y_t^2;t=0,1,..., T\}$ at lags $0$, $1$ and $2$.  

For the Bayesian optimization approach to learning the weights, 50 weight vectors were generated initially
according to a maximin latin hypercube design, with the value $\omega_1=1$ being fixed.   
Using a stationary radial basis kernel and constant mean for the Gaussian process surrogate, 
$100$ further steps were conducted, 
with an expected improvement (EI) acquisition function and covariance hyperparameters re-estimated at every iteration. 
For the linear pooling approach, we combine forecast densities from
3 different summary statistic choices, based on the auxiliary model
approach for the scoring rules LS, CLS$_{20}$ and CLS$_{80}$, and evaluate
the ensemble in terms of each of these scoring rules.  

Tables \ref{auxiliarysslag1}-\ref{autocovsslag2} show the results of the simulation study.  We make
four observations.  First, adaptive distance methods generally outperform fixed distance methods
across different posterior samplers (ABC or TabPFN), different lead times and summary statistic
choices.  Second, the gains are larger for the adaptive distance method when the autocovariance
summary statistic are used;  this makes sense, since if the summary statistics are already
chosen to achieve good performance in terms of the given scoring rule, then the additional
benefit of adaptive distance learning might be smaller.  Third, TabPFN generally outperforms
the corresponding method using ABC for the autocovariance summary statistics,
and this might suggest that with higher-dimensional summary statistics TabPFN is
preferred.  Finally, for the auxiliary summary statistics, 
linear pooling is the best approach overall in terms of the log, CLS$_{10}$ and
CLS$_{20}$ scores for lag one forecasts, and better for all scoring rules for the lag 2 forecast horizon when TabPFN is used.

\begin{table}[H]
    \centering
    \caption{\label{auxiliarysslag1}Scoring rule evaluations of predictive performance for one-step ahead forecasts 
    with auxiliary-model summary statistics and using ABC and TabPFN for generating posterior samples.
    Higher values are better, and the best value in each column is indicated in bold.} 
    \sisetup{detect-weight=true, detect-family=true, table-format=-1.4}
    \setlength{\tabcolsep}{3pt}
    \begin{tabular}{l *{7}{S}} 
        \hline
       & {LS} & {$CLS_{10}$} & {$CLS_{20}$} & {$CLS_{80}$} & {$CLS_{90}$} & {negIS} & {negCRPS} \\ 
        \hline
        \multicolumn{8}{c}{ABC} \\
        \hline
        Adaptive distance &  {-1.3392 }& { -0.3811} & \bfseries{-0.6145} & \bfseries{ -0.4461} & \bfseries{ -0.2629}&   \bfseries{ -4.2361} & \bfseries{ -0.5297}\\
        Fixed distance    & {-1.3550}
 & {-0.3981}
&{-0.6259} &{ -0.4515} &  {-0.2884} &{ -4.2610 } & { -0.5331} \\
        Linear pooling    & \bfseries{ -1.3185} &  \bfseries{ -0.3772}& {  -0.6164}&{ -0.4527} & {-0.2632} & {   -4.6011}& {-0.5304} \\
        \hline
        \multicolumn{8}{c}{TabPFN} \\
        \hline
        Adaptive distance &  { -1.3260}& {-0.4055} & { -0.6107} & { -0.4435} & {   -0.2624} &{-4.2959} & \bfseries{    -0.5282} \\
        Fixed distance    & {   -1.3280} & {-0.4073} & {   -0.6112} & \bfseries{-0.4433} & { -0.2630} & \bfseries{-4.2801} & {  -0.5284}\\
        Linear pooling    & \bfseries{-1.3055}& \bfseries{-0.3715} & \bfseries{ -0.6059} & { -0.4434} & \bfseries{-0.2582} & {-4.4306} & {-0.5285} \\
        \hline
    \end{tabular}
\end{table}

\begin{table}[H]
    \centering
    \caption{\label{autocovsslag1}Scoring rule evaluations of predictive performance for one-step ahead forecasts 
    with autocovariance summary statistics and using ABC and TabPFN for generating posterior samples.
    Higher values are better, and the best value in each column is indicated in bold.} 
    \sisetup{detect-weight=true, detect-family=true, table-format=-1.4}
    \setlength{\tabcolsep}{3pt}
    \begin{tabular}{l *{7}{S}} 
        \hline
       & {LS} & {$CLS_{10}$} & {$CLS_{20}$} & {$CLS_{80}$} & {$CLS_{90}$} & {negIS} & {negCRPS} \\ 
        \hline
        \multicolumn{8}{c}{ABC} \\
        \hline
        Adaptive distance & \bfseries{ -1.4380}& \bfseries{ -0.3991}& {   -0.6265} &\bfseries{ -0.5612 } & \bfseries{ -0.3676}&  \bfseries{ -5.0683} & \bfseries{ -0.5456} \\
        Fixed distance    & { -1.5039} &  { -0.4571} & \bfseries{  -0.6410} & { -0.5889} & { -0.3857} & {-5.8871} &  { -0.5774}\\
        \hline
        \multicolumn{8}{c}{TabPFN} \\
        \hline
        Adaptive distance &\bfseries{ -1.3223} & {-0.3757} & \bfseries{-0.6075}  & \bfseries{-0.4677} & \bfseries{ -0.2752} & \bfseries{-4.5884} & \bfseries{-0.5291} \\
        Fixed distance    &  { -1.3226} & \bfseries{-0.3738}&  {-0.6085} & {-0.4687} &{  -0.2790}  &{   -4.6355} &{-0.5296} \\
        \hline
    \end{tabular}
\end{table}

\begin{table}[H]
    \centering
    \caption{\label{auxiliarysslag2}Scoring rule evaluations of predictive performance for two-step ahead forecasts 
    with auxiliary model summary statistics and using ABC and TabPFN for generating posterior samples.
    Higher values are better, and the best value in each column is indicated in bold.} 
    \sisetup{detect-weight=true, detect-family=true, table-format=-1.4}
    \setlength{\tabcolsep}{3pt}
    \begin{tabular}{l *{7}{S}} 
        \hline
       & {LS} & {$CLS_{10}$} & {$CLS_{20}$} & {$CLS_{80}$} & {$CLS_{90}$} & {negIS} & {negCRPS} \\ 
        \hline
        \multicolumn{8}{c}{ABC} \\
        \hline
        Adaptive distance & { -1.2944} & \bfseries{-0.3606}  &  \bfseries{-0.5893} & \bfseries{-0.4477} & {-0.2666} &  \bfseries{-3.8977} & \bfseries{  -0.5147} \\
        Fixed distance    &  { -1.3189}&  {-0.3846} & { -0.6060} &  {-0.4509}&  {-0.2840}& { -4.0567}  &  {-0.5279} \\
        Linear pooling    & \bfseries{-1.2921}&{ -0.3648
} & {-0.5971} & 
       { -0.4506} & \bfseries{-0.2619} & {-4.3385} & { -0.5231}\\
        \hline
        \multicolumn{8}{c}{TabPFN} \\
        \hline
        Adaptive distance & { -1.2496 }& {-0.3747}& {  -0.5713} & {-0.4449} &{ -0.2645}  &{  -4.1002
} & { -0.5109}\\
        Fixed distance    & { -1.2547} &  { -0.3903}&  {-0.5798 } & { -0.4450 } &  { -0.2650} &  { -4.0766}&  {  -0.5128}\\
        Linear pooling    &  \bfseries{  -1.2446} & \bfseries{-0.3374} & \bfseries{ -0.5624} &  \bfseries{ -0.4446} &  \bfseries{-0.2593} & \bfseries{ -3.9080} &   \bfseries{ -0.5162}\\
        \hline
    \end{tabular}
\end{table}

\begin{table}[H]
    \centering
    \caption{\label{autocovsslag2}Scoring rule evaluations of predictive performance for two-step ahead forecasts 
    with autocovariance summary statistics and using ABC and TabPFN for generating posterior samples.
    Higher values are better, and the best value in each column is indicated in bold.} 
    \sisetup{detect-weight=true, detect-family=true, table-format=-1.4}
    \setlength{\tabcolsep}{3pt}
    \begin{tabular}{l *{7}{S}} 
        \hline
       & {LS} & {$CLS_{10}$} & {$CLS_{20}$} & {$CLS_{80}$} & {$CLS_{90}$} & {negIS} & {negCRPS} \\ 
        \hline
        \multicolumn{8}{c}{ABC} \\
        \hline
        Adaptive distance & \bfseries{-1.3867}  & \bfseries{-0.3800}& \bfseries{-0.6374} & \bfseries{-0.4904} & \bfseries{ -0.3294} & \bfseries{ -4.9551} & \bfseries{  -0.5344} \\
        Fixed distance    & { -1.4555}  &{-0.3953}  &  { -0.6384}&{-0.5421} &  {-0.3708} & {-5.7367} &   {-0.5655} \\
        \hline
        \multicolumn{8}{c}{TabPFN} \\
        \hline
        Adaptive distance & { -1.2672}  & {-0.3427} &{-0.5675}  &  \bfseries{   -0.4604}&{-0.2740}  &  \bfseries{ -4.0648}& \bfseries{ -0.5185}  \\
        Fixed distance    & \bfseries{-1.2724} & \bfseries{-0.3421} & \bfseries{-0.5741} &{ -0.4606} &\bfseries{-0.2728}  &  {-4.0858}& {-0.5195} \\
        \hline
    \end{tabular}
\end{table}

\subsection{Real data example:  stochastic volatility model with an intractable
transition density}

Next, we consider another example from \cite{weerasinghe+lmf23}.  It considers
forecasting for a real data set using a stochastic volatility model with
an intractable transition density involving increments from a 
heavy-tailed $\alpha$-stable distribution.  Although computation of the
$\alpha$-stable distribution is intractable, it is possible to simulate
from the model, so that likelihood-free methods are attractive for models
involving the $\alpha$-stable distribution \citep{peters+sf12}.  
The model is defined as follows:
\begin{align}
  y_t & = \exp(z_t/2)\epsilon_y, \;\;\;\epsilon_y\sim N(0,1),  \label{obseq-real} \\
  z_t & = \mu_z+\phi z_{t-1}+\sigma_z \epsilon_z, \;\;\;\epsilon_z\sim S(\alpha,\tau,\mu,\sigma),  \label{state-eq-real}
\end{align}
where $S(\alpha,\tau,\mu,\sigma)$ denotes the $\alpha$-stable
distribution with tail index $\alpha\in [1,2]$, 
skewness parameter $\tau$, location $\mu$ and scale $\sigma$.  
Following \cite{weerasinghe+lmf23} we fix $\mu=0$, $\tau=-1$ and $\sigma=1$ 
but treat $\tau$ as part of the unknowns, which are $\theta_a=(\mu_z,\phi,\sigma_z,\tau)^\top$.  We use the same priors as 
\cite{weerasinghe+lmf23}, with $\mu_z\sim U[-1,1]$, $\phi\sim U[0.5,0.99]$, 
$\sigma_z\sim U[0,0.3]$ and $\tau\sim U[1,2]$. 

The real data consists of close-to-close daily returns on the S\&P500 index
from 4 January 2010 to 31 December 2019.   There are 2516 observations, 
and we split into training, validation and forecasting periods using
$T=1016$, $\widetilde{T}=1000$ and $\breve{T}=500$.  Following
\cite{weerasinghe+lmf23}, we consider auxiliary summary statistics based
on a GARCH(1,1) model, as well as the 6-dimensional autocovariance
summary statistics considered in the simulation study and assess the forecasting performance using
the censored log scoring rules LS, CLS$_{10}$, CLS$_{20}$, CLS$_{80}$ and CLS$_{90}$.  

Tables \ref{auxiliarysslag1real} and \ref{autocovsslag1real} show 
the results for lag 1 forecasts for the auxiliary summary statistics and 
autocovariance summary statistics respectively.  We make three observations.  
First, in contrast to the simulation study, there is little difference
in performance between ABC and TabPFN in this example.  
Second, the adaptive distance approach outperforms fixed distance.  
Third, linear pooling performs best for the log score, but is generally
inferior to both adaptive and fixed distance methods for other scoring
rules.

\begin{table}[htbp]
    \centering
    \caption{\label{auxiliarysslag1real}Scoring rule evaluations of predictive performance for one-step ahead forecasts 
    with auxiliary-model summary statistics and using ABC and TabPFN for generating posterior samples for the S\&P 500 data. 
    Higher values are better, and the best value in each column is indicated in bold.} 
    \sisetup{detect-weight=true, detect-family=true}
    \setlength{\tabcolsep}{4pt}
    \begin{tabular}{l S[table-format=1.4(1.4)] *{4}{S[table-format=1.4]}} 
        \hline
       & {LS} & {$CLS_{10}$} & {$CLS_{20}$} & {$CLS_{80}$} & {$CLS_{90}$} \\ 
        \hline
        \multicolumn{6}{c}{ABC} \\
        \hline
        Adaptive distance & { 3.3854}  & \bfseries{0.0471} & \bfseries{0.2454}  &  \bfseries{ 0.3887}  &  { 0.0695}\\
        Fixed distance    & { 3.3767}  & { 0.0463} &  { 0.2451}& {  0.3853}&  \bfseries{  0.0721}  \\
        Linear pooling    &  \bfseries{3.3939}  & {0.0402}  & {  0.2389} & { 0.3831}  & {0.0677}  \\
        \hline
        \multicolumn{6}{c}{TabPFN} \\
        \hline
        Adaptive distance &{  3.3873}  &{0.0455}  & \bfseries{0.2466} & \bfseries{ 0.3889} &  \bfseries{ 0.0763}\\
        Fixed distance    & {3.3853} & \bfseries{0.0459} & { 0.2460} &{ 0.3864}  & {0.0743} \\
        Linear pooling    & \bfseries{3.3950} & { 0.0430} & { 0.2424} & { 0.3819} & { 0.0736} \\
        \hline
    \end{tabular}
\end{table}

\begin{table}[htbp]
    \centering
    \caption{\label{autocovsslag1real}Scoring rule evaluations of predictive performance for one-step ahead forecasts 
    with autocovariance summary statistics and using ABC and TabPFN for generating posterior samples for the S\&P 500 data. 
    Higher values are better, and the best value in each column is indicated in bold.} 
    \sisetup{detect-weight=true, detect-family=true, table-format=1.4}
    \setlength{\tabcolsep}{4pt}
    \begin{tabular}{l *{5}{S}} 
        \hline
       & {LS} & {$CLS_{10}$} & {$CLS_{20}$} & {$CLS_{80}$} & {$CLS_{90}$} \\ 
        \hline
        \multicolumn{6}{c}{ABC} \\
        \hline
        Adaptive distance &  \bfseries{3.3916} &  \bfseries{0.0469} &  \bfseries{ 0.2507 } &  \bfseries{ 0.3793 } &  \bfseries{ 0.0625} \\
        Fixed distance    & {3.3313} & {0.0424}  &  {0.2477}  &  {0.3537} &   { 0.0701}\\
        \hline
        \multicolumn{6}{c}{TabPFN} \\
        \hline
        Adaptive distance & \bfseries{ 3.4041} & \bfseries{  0.0464} & { 0.2493} & \bfseries{ 0.3861} & \bfseries{0.0724} \\
        Fixed distance    & {3.4034} & {0.0462} & \bfseries{ 0.2500}&{0.3829} & { 0.0699} \\
        \hline
    \end{tabular}
\end{table}

\section{Discussion}\label{sec:Discussion}
 
Adaptive distance learning in ABC has received considerable attention, 
but not in the context of optimizing forecasting performance for misspecified time series. 
Our work addresses this, showing that adaptive distance learning can improve forecasting
performance under a chosen scoring rule, both for ABC and for TabPFN with localization. 
We have also framed the use of linear opinion pools as corresponding to a randomized choice 
of distance, connecting them to the pooled LFI posteriors of \cite{frazier+dkn25}. 
The summary statistics of \cite{weerasinghe+lmf23}, which are based on diverse scoring rules, 
provide an effective and natural way to specify the pool members combined for forecasting.

Interest in learning posterior distributions that maximize predictive performance in 
Bayesian models has increased recently, extending well beyond time series. Notable examples 
include predictive variational inference (PVI) and predictively-oriented (PrO) posteriors
\citep{lai+ly25,mclatchie+cfk20}, with \cite{lai+ly25} 
extending the PVI approach to SBI applications. The PVI and PrO posterior distributions 
need not concentrate to a single point asymptotically under misspecification, and our 
adaptive distance posteriors share this property when the distance is optimized for 
forecasting. This is different to other generalized Bayesian approaches, and can be beneficial
for prediction under misspecification, since concentration of the posterior to a point results in plug-in
prediction using the wrong model, which could be undesirable.  
\cite{shen+do26} is a pioneering work on predictively-oriented Kalman filtering for non-linear time series, and further development of PVI posteriors for SBI, building on \cite{lai+ly25}, is an attractive direction for future work. While the posteriors obtained through 
distance learning alone have limited expressiveness, we believe distance learning 
can contribute to the development of more flexible Bayesian approaches for 
forecasting in misspecified time series.

\section*{Acknowledgements}

David Nott's research was supported by the Ministry of Education, Singapore, under the Academic Research Fund Tier 2 (MOE-T2EP20123-0009).  The authors thank
Chaya Weerasinghe for sharing her code.

\bibliographystyle{apalike}
\addcontentsline{toc}{section}{\refname}
\bibliography{references}

@article{shen+do26,
  title={Predictively-Oriented {K}alman Filtering},
  author={Shen, Zheyang and Duran-Martin, Gerardo and Oates, Chris},
  journal={arXiv preprint arXiv:2606.03230},
  year={2026}
}

@article{kelly+fwd26,
  title={Preconditioned robust neural posterior estimation for misspecified simulators},
  author={Kelly, Ryan P and Frazier, David T and Warne, David J and Drovandi, Christopher C},
  journal={arXiv preprint arXiv:2602.18004},
  year={2026}
}

@article{aicher+pnff25,
author = {Christopher Aicher and Srshti Putcha and Christopher Nemeth and Paul Fearnhead and Emily Fox},
title = {{Stochastic Gradient MCMC for Nonlinear State Space Models}},
volume = {20},
journal = {Bayesian Analysis},
number = {1},
pages = {83 -- 105},
year = {2025}}

@inproceedings{gordon+ss93,
  title={Novel approach to nonlinear/non-{G}aussian {B}ayesian state estimation},
  author={Gordon, Neil J and Salmond, David J and Smith, Adrian FM},
  booktitle={IEE proceedings F (radar and signal processing)},
  volume={140, no. 2},
  pages={107--113},
  year={1993},
  organization={IET}
}

@article{peters+sf12,
    title = {Likelihood-free {B}ayesian inference for $\alpha$-stable models},
    journal = {Computational Statistics \& Data Analysis},
    volume = {56},
    number = {11},
    pages = {3743-3756},
    year = {2012},
    author = {G.W. Peters and S.A. Sisson and Y. Fan}}

@article{beaumont+zb02,
	author = {Beaumont, M. A. and Zhang, W. and Balding, D. J.},
	journal = {Genetics},
	number = {4},
	pages = {2025--2035},
	title = {{Approximate Bayesian computation in population genetics}},
	volume = {162},
	year = {2002}}

@article{frazier+dkn25,
	author = {David T. Frazier and Christopher Drovandi and Lucas Kock and David J. Nott},
	journal = {Bayesian Analysis},
	pages = {1 -- 25},
	title = {{Pooling Information in Likelihood-Free Inference}},
	year = {2025}}

@article{miller+d19,
	author = {Jeffrey W. Miller and David B. Dunson},
	journal = {Journal of the American Statistical Association},
	number = {527},
	pages = {1113--1125},
	title = {Robust {B}ayesian Inference via Coarsening},
	volume = {114},
	year = {2019}}

@article{stock+w04,
	author = {Stock, James H. and Watson, Mark W.},
	journal = {Journal of Forecasting},
	number = {6},
	pages = {405-430},
	title = {Combination forecasts of output growth in a seven-country data set},
	volume = {23},
	year = {2004}}

@article{stone1961,
	author = {Stone, Mervyn},
	journal = {The Annals of Mathematical Statistics},
	pages = {1339--1342},
	title = {The opinion pool},
	year = {1961}}

@article{wang+hfk23,
	author = {Xiaoqian Wang and Rob J. Hyndman and Feng Li and Yanfei Kang},
	journal = {International Journal of Forecasting},
	number = {4},
	pages = {1518-1547},
	title = {Forecast combinations: An over 50-year review},
	volume = {39},
	year = {2023}}

@inproceedings{yao+bd24,
	author = {Yao, Yuling and R\'{e}galdo-Saint Blancard, Bruno and Domke, Justin},
	booktitle = {Proceedings of The 27th International Conference on Artificial Intelligence and Statistics},
	editor = {Dasgupta, Sanjoy and Mandt, Stephan and Li, Yingzhen},
	pages = {4267--4275},
	publisher = {PMLR},
	series = {Proceedings of Machine Learning Research},
	title = {Simulation-Based Stacking},
	volume = {238},
	year = {2024}}

@article{yao+vsg18,
	author = {Yuling Yao and Aki Vehtari and Daniel Simpson and Andrew Gelman},
	journal = {Bayesian Analysis},
	number = {3},
	pages = {917 -- 1007},
	title = {Using Stacking to Average {B}ayesian Predictive Distributions (with Discussion)},
	volume = {13},
	year = {2018}}

@book{sisson+fb18,
	editor = {S. A. Sisson and Y. Fan and M. A. Beaumont},
	publisher = {Chapman {\&} Hall/CRC},
	title = {Handbook of Approximate Bayesian Computation},
	year = {2018}}

@inproceedings{nagler23,
	author = {Nagler, Thomas},
	booktitle = {Proceedings of the 40th International Conference on Machine Learning},
	editor = {Krause, Andreas and Brunskill, Emma and Cho, Kyunghyun and Engelhardt, Barbara and Sabato, Sivan and Scarlett, Jonathan},
	pages = {25660--25676},
	publisher = {PMLR},
	series = {Proceedings of Machine Learning Research},
	title = {Statistical Foundations of Prior-Data Fitted Networks},
	volume = {202},
	year = {2023}}

@misc{grinsztajn25,
      title={{TabPFN-2.5}: Advancing the State of the Art in Tabular Foundation Models}, 
      author={Léo Grinsztajn and Klemens Flöge and Oscar Key and Felix Birkel and Philipp Jund and Brendan Roof and Benjamin Jäger and Dominik Safaric and Simone Alessi and Adrian Hayler and Mihir Manium and Rosen Yu and Felix Jablonski and Shi Bin Hoo and Anurag Garg and Jake Robertson and Magnus Bühler and Vladyslav Moroshan and Lennart Purucker and Clara Cornu and Lilly Charlotte Wehrhahn and Alessandro Bonetto and Bernhard Schölkopf and Sauraj Gambhir and Noah Hollmann and Frank Hutter},
      year={2025},
      eprint={2511.08667},
      primaryClass={cs.LG},
      url={https://arxiv.org/abs/2511.08667}, 
}

@article{lai+ly25,
	archiveprefix = {arXiv},
	author = {Jinlin Lai and Antonio Linero and Yuling Yao},
    journal= {arXiv preprint: arXiv2410.14843},
	title = {Predictive variational inference: Learn the predictively optimal posterior distribution},
	year = {2025}
}

@article{mclatchie+cfk20,
	author = {Yann McLatchie and Badr-Eddine Cherief-Abdellatif and David T. Frazier and Jeremias Knoblauch},
	journal = {arXiv preprint arXiv:2510.01915},
	title = {Predictively Oriented Posteriors},
	year = {2025}}

@inproceedings{vaswani+spujgkp17,
	author = {Vaswani, Ashish and Shazeer, Noam and Parmar, Niki and Uszkoreit, Jakob and Jones, Llion and Gomez, Aidan N and Kaiser, {\L}ukasz and Polosukhin, Illia},
	booktitle = {Advances in Neural Information Processing Systems},
	editor = {I. Guyon and U. Von Luxburg and S. Bengio and H. Wallach and R. Fergus and S. Vishwanathan and R. Garnett},
	publisher = {Curran Associates, Inc.},
	title = {Attention is All you Need},
	volume = {30},
	year = {2017}}

@book{peters+js17,
	author = {Peters, Jonas and Janzing, Dominik and Sch{\"o}lkopf, Bernhard},
	publisher = {The MIT Press},
	title = {Elements of causal inference: foundations and learning algorithms},
	year = {2017}}

@article{zhang+ttl25,
	author = {Zhang, Qiong and Tan, Yan Shuo and Tian, Qinglong and Li, Pengfei},
	journal = {arXiv preprint arXiv:2505.20003},
	title = {{TabPFN}: One Model to Rule Them All?},
	year = {2025}}

@article{hollman+mpkkhsh25,
	author = {Hollmann, Noah and M{\"u}ller, Samuel and Purucker, Lennart and Krishnakumar, Arjun and K{\"o}rfer, Max and Hoo, Shi Bin and Schirrmeister, Robin Tibor and Hutter, Frank},
	journal = {Nature},
	number = {8045},
	pages = {319--326},
	title = {Accurate predictions on small data with a tabular foundation model},
	volume = {637},
	year = {2025}}

@article{frazier+rr20,
	author = {Frazier, David T. and Robert, Christian P. and Rousseau, Judith},
	journal = {Journal of the Royal Statistical Society: Series B (Statistical Methodology)},
	number = {2},
	pages = {421--444},
	title = {Model misspecification in approximate {B}ayesian computation: consequences and diagnostics},
	volume = {82},
	year = {2020}}

@article{vetter+ggm25,
	author = {Vetter, Julius and Gloeckler, Manuel and Gedon, Daniel and Macke, Jakob H},
	journal = {arXiv preprint arXiv:2504.17660},
	title = {Effortless, Simulation-Efficient {B}ayesian Inference using Tabular Foundation Models},
	year = {2025}}

@article{canale+r16,
	author = {Antonio Canale and Matteo Ruggiero},
	journal = {Electronic Journal of Statistics},
	month = {1},
	number = {2},
	pages = {3265--3286},
	title = {Bayesian nonparametric forecasting of monotonic functional time series},
	volume = {10},
	year = {2016}}

@article{mckinley+cd09,
	author = {Trevelyan McKinley and Alex R Cook and Robert Deardon},
	journal = {The International Journal of Biostatistics},
	number = {1},
	title = {Inference in epidemic models without likelihods},
	volume = {5},
	year = {2009}}

@article{jarvenpaa+c23,
	author = {J{\"a}rvenp{\"a}{\"a}, Marko and Corander, Jukka},
	journal = {Statistics and Computing},
	number = {2},
	pages = {42},
	title = {On predictive inference for intractable models via approximate {B}ayesian computation},
	volume = {33},
	year = {2023}}

@article{jasra+smm12,
	author = {Jasra, Ajay and Singh, Sumeetpal S. and Martin, James S. and McCoy, Emma},
	journal = {Statistics and Computing},
	number = {6},
	pages = {1223--1237},
	title = {Filtering via approximate {B}ayesian computation},
	volume = {22},
	year = {2012}}

@article{frazier+lmk25,
	author = {Frazier, David T. and Loaiza-Maya, Rub{\'e}n and Martin, Gael M. and Koo, Bonsoo},
	journal = {Journal of Computational and Graphical Statistics},
	number = {1},
	pages = {84--95},
	title = {Loss-Based Variational {B}ayes Prediction},
	volume = {34},
	year = {2025},
	year1 = {2025}}

@article{weerasinghe+lmf23,
	author = {Weerasinghe, Chaya and Loaiza-Maya, Rub{\'e}n and Martin, Gael M. and Frazier, David T.},
	journal = {International Journal of Forecasting},
	number = {1},
	pages = {270--289},
	title = {{ABC}-based forecasting in misspecified state space models},
	volume = {41},
	year = {2025}}

@article{legramanti+da25,
	author = {Legramanti, Sirio and Durante, Daniele and Alquier, Pierre},
	journal = {The Annals of Statistics},
	number = {1},
	pages = {37--60},
	title = {Concentration of discrepancy-based approximate {B}ayesian computation via {R}ademacher complexity},
	volume = {53},
	year = {2025}}

@article{frazier20,
	author = {Frazier, David T},
	journal = {arXiv preprint arXiv:2006.14126},
	title = {Robust and efficient approximate {B}ayesian computation: A minimum distance approach},
	year = {2020}}

@inproceedings{park+js16,
	author = {Park, Mijung and Jitkrittum, Wittawat and Sejdinovic, Dino},
	booktitle = {Proceedings of the 19th International Conference on Artificial Intelligence and Statistics},
	editor = {Gretton, Arthur and Robert, Christian C.},
	pages = {398--407},
	publisher = {PMLR},
	series = {Proceedings of Machine Learning Research},
	title = {K2-ABC: Approximate {B}ayesian Computation with Kernel Embeddings},
	volume = {51},
	year = {2016}}

@article{nguyen+alf20,
	author = {Nguyen, Hien Duy and Arbel, Julyan and L{\"u}, Hongliang and Forbes, Florence},
	doi = {10.1109/ACCESS.2020.3009878},
	journal = {IEEE Access},
	pages = {131683-131698},
	title = {Approximate {B}ayesian Computation Via the Energy Statistic},
	volume = {8},
	year = {2020}}

@article{bernton19,
	author = {Bernton, Espen and Jacob, Pierre E. and Gerber, Mathieu and Robert, Christian P.},
	journal = {Journal of the Royal Statistical Society Series B: Statistical Methodology},
	number = {2},
	pages = {235--269},
	title = {Approximate {B}ayesian Computation with the {W}asserstein Distance},
	url = {https://doi.org/10.1111/rssb.12312},
	volume = {81},
	year = {2019}}

@inproceedings{jiang18,
	author = {Jiang, Bai},
	booktitle = {Proceedings of the Twenty-First International Conference on Artificial Intelligence and Statistics},
	editor = {Storkey, Amos and Perez-Cruz, Fernando},
	pages = {1711--1721},
	publisher = {PMLR},
	series = {Proceedings of Machine Learning Research},
	title = {Approximate {B}ayesian Computation with {K}ullback-{L}eibler Divergence as Data Discrepancy},
	volume = {84},
	year = {2018}}

@article{drovandi+f22,
	author = {Drovandi, Christopher and Frazier, David T.},
	date = {2022/05/19},
	doi = {10.1007/s11222-022-10092-4},
	id = {Drovandi2022},
	isbn = {1573-1375},
	journal = {Statistics and Computing},
	number = {3},
	pages = {42},
	title = {A comparison of likelihood-free methods with and without summary statistics},
	url = {https://doi.org/10.1007/s11222-022-10092-4},
	volume = {32},
	year = {2022}}

@article{thomas+slkcp25,
	author = {Thomas, Owen and S{\'a}-Le{\~a}o, Raquel and de Lencastre, Herm{\'\i}nia and Kaski, Samuel and Corander, Jukka and Pesonen, Henri},
	journal = {Computational Statistics},
	number = {8},
	pages = {4399--4439},
	title = {Misspecification-robust likelihood-free inference in high dimensions},
	volume = {40},
	year = {2025}}

@article{schalte+h23,
	author = {Sch{\"a}lte, Yannik and Hasenauer, Jan},
	journal = {Plos one},
	number = {5},
	pages = {e0285836},
	publisher = {Public Library of Science San Francisco, CA USA},
	title = {Informative and adaptive distances and summary statistics in sequential approximate {B}ayesian computation},
	volume = {18},
	year = {2023}}

@article{schalte+ah21,
	author = {Sch{\"a}lte, Yannik and Alamoudi, Emad and Hasenauer, Jan},
	journal = {bioRxiv},
	publisher = {Cold Spring Harbor Laboratory},
	title = {Robust adaptive distance functions for approximate {B}ayesian inference on outlier-corrupted data},
	volume = {2021.07.29.454327},
	year = {2021}}

@article{gutmann+c16,
	author = {Michael U. Gutmann and Jukka Corander},
	journal = {Journal of Machine Learning Research},
	number = {125},
	pages = {1--47},
	title = {Bayesian Optimization for Likelihood-Free Inference of Simulator-Based Statistical Models},
	url = {http://jmlr.org/papers/v17/15-017.html},
	volume = {17},
	year = {2016}}

@article{harrison+b20,
	address = {Mathematical Institute, Mathematical Sciences Building, University of Warwick, Coventry, United Kingdom.; Mathematical Institute, Andrew Wiles Building, University of Oxford, Oxford, United Kingdom.},
	auid = {ORCID: 0000-0002-2748-9921},
	author = {Harrison, Jonathan U and Baker, Ruth E},
	cois = {The authors have declared that no competing interests exist.},
	crdt = {2020/08/08 06:00},
	date = {2020},
	dcom = {20201005},
	dep = {20200806},
	doi = {10.1371/journal.pone.0236954},
	edat = {2020/08/08 06:00},
	gr = {BB/R000816/1/BB{\_}/Biotechnology and Biological Sciences Research Council/United Kingdom; BB/R00816/1/BB{\_}/Biotechnology and Biological Sciences Research Council/United Kingdom},
	issn = {1932-6203 (Electronic); 1932-6203 (Linking)},
	jid = {101285081},
	journal = {PLoS One},
	jt = {PloS one},
	language = {eng},
	lid = {10.1371/journal.pone.0236954 {$[$}doi{$]$}; e0236954},
	lr = {20250719},
	mh = {*Algorithms; *Bayes Theorem; Biochemical Phenomena; Biometry/*methods; Computer Simulation; Likelihood Functions; Markov Chains; Metabolic Networks and Pathways; Models, Biological; Models, Statistical; Monte Carlo Method; Regression Analysis; Stochastic Processes},
	mhda = {2020/10/06 06:00},
	number = {8},
	own = {NLM},
	pages = {e0236954},
	phst = {2020/03/02 00:00 {$[$}received{$]$}; 2020/07/16 00:00 {$[$}accepted{$]$}; 2020/08/08 06:00 {$[$}entrez{$]$}; 2020/08/08 06:00 {$[$}pubmed{$]$}; 2020/10/06 06:00 {$[$}medline{$]$}; 2020/08/06 00:00 {$[$}pmc-release{$]$}},
	pii = {PONE-D-20-06109},
	pl = {United States},
	pmc = {PMC7410215},
	pmcr = {2020/08/06},
	pmid = {32760106},
	pst = {epublish},
	pt = {Journal Article; Research Support, Non-U.S. Gov't},
	sb = {IM},
	status = {MEDLINE},
	title = {An automatic adaptive method to combine summary statistics in approximate {B}ayesian computation.},
	volume = {15},
	year = {2020}}

@article{gutmann+dkc18,
	author = {Gutmann, Michael U. and Dutta, Ritabrata and Kaski, Samuel and Corander, Jukka},
	date = {2018/03/01},
	doi = {10.1007/s11222-017-9738-6},
	id = {Gutmann2018},
	isbn = {1573-1375},
	journal = {Statistics and Computing},
	number = {2},
	pages = {411--425},
	title = {Likelihood-free inference via classification},
	url = {https://doi.org/10.1007/s11222-017-9738-6},
	volume = {28},
	year = {2018}}

@article{jung+m11,
	author = {Hsuan Jung and Paul Marjoram},
	doi = {doi:10.2202/1544-6115.1586},
	journal = {Statistical Applications in Genetics and Molecular Biology},
	number = {1},
	pages = {45},
	title = {Choice of Summary Statistic Weights in Approximate {B}ayesian Computation},
	url = {https://doi.org/10.2202/1544-6115.1586},
	volume = {10},
	year = {2011}}

@article{schmon+ck20,
	author = {Schmon, Sebastian M and Cannon, Patrick W and Knoblauch, Jeremias},
	journal = {arXiv preprint arXiv:2011.08644},
	title = {Generalized posteriors in approximate {B}ayesian computation},
	year = {2020}}

@article{kelly+nfwd24,
	author = {Ryan P. Kelly and David J Nott and David Tyler Frazier and David J Warne and Christopher Drovandi},
	issn = {2835-8856},
	journal = {Transactions on Machine Learning Research},
	title = {Misspecification-robust Sequential Neural Likelihood for Simulation-based Inference},
	volume = {\url{https://openreview.net/forum?id=tbOYJwXhcY}},
	year = {2024}}

@article{frazier+d21,
	author = {David T. Frazier and Christopher Drovandi},
	journal = {Journal of Computational and Graphical Statistics},
	number = {4},
	pages = {958--976},
	title = {Robust Approximate {B}ayesian Inference With Synthetic Likelihood},
	volume = {30},
	year = {2021}}

@article{ward2022robust,
	author = {Ward, Daniel and Cannon, Patrick and Beaumont, Mark and Fasiolo, Matteo and Schmon, Sebastian M},
	journal = {arXiv preprint arXiv:2210.06564},
	title = {Robust Neural Posterior Estimation and Statistical Model Criticism},
	year = {2022}}

@article{wilkinson13,
	author = {Richard David Wilkinson},
	journal = {Statistical Applications in Genetics and Molecular Biology},
	number = {2},
	pages = {129 - 141},
	title = {Approximate {B}ayesian computation ({ABC}) gives exact results under the assumption of model error},
	volume = {12},
	year = {2013}}

@inproceedings{balandat20,
	author = {Balandat, Maximilian and Karrer, Brian and Jiang, Daniel and Daulton, Samuel and Letham, Ben and Wilson, Andrew G and Bakshy, Eytan},
	booktitle = {Advances in Neural Information Processing Systems},
	editor = {H. Larochelle and M. Ranzato and R. Hadsell and M.F. Balcan and H. Lin},
	pages = {21524--21538},
	publisher = {Curran Associates, Inc.},
	title = {BoTorch: A Framework for Efficient {Monte-Carlo Bayesian} Optimization},
	url = {https://proceedings.neurips.cc/paper_files/paper/2020/file/f5b1b89d98b7286673128a5fb112cb9a-Paper.pdf},
	volume = {33},
	year = {2020}}

@book{garnett23,
	author = {Garnett, Roman},
	doi = {10.1017/9781108348973},
	place = {Cambridge},
	publisher = {Cambridge University Press},
	title = {Bayesian Optimization},
	year = {2023}}

@article{prangle17,
	author = {Dennis Prangle},
	journal = {Bayesian Analysis},
	number = {1},
	pages = {289 -- 309},
	title = {{Adapting the ABC Distance Function}},
	volume = {12},
	year = {2017}}

@article{martin+mfmr19,
	author = {Martin, Gael M. and McCabe, Brendan P. M. and Frazier, David T. and Maneesoonthorn, Worapree and Robert, Christian P.},
	journal = {Journal of Computational and Graphical Statistics},
	number = {3},
	pages = {508-522},
	title = {Auxiliary Likelihood-Based Approximate {B}ayesian Computation in State Space Models},
	volume = {28},
	year = {2019}}

@misc{gleim+p13,
	author = {Gleim, A. and Pigorsch, C.},
	howpublished = {Technical Report, University of Bonn},
	title = {Approximate {B}ayesian computation with indirect summary statistics},
	year = {2013}}

@article{drovandi+pl15,
	author = {Christopher C. Drovandi and Anthony N. Pettitt and Anthony Lee},
	journal = {Statistical Science},
	number = {1},
	pages = {72 -- 95},
	title = {Bayesian Indirect Inference Using a Parametric Auxiliary Model},
	volume = {30},
	year = {2015}}

@article{dunsmore68,
	author = {Dunsmore, I. R.},
	journal = {Journal of the Royal Statistical Society: Series B},
	number = {2},
	pages = {396-405},
	title = {A {B}ayesian Approach to Calibration},
	volume = {30},
	year = {1968}}

@article{diks+pv11,
	author = {Cees Diks and Valentyn Panchenko and Dick {van Dijk}},
	journal = {Journal of Econometrics},
	number = {2},
	pages = {215-230},
	title = {Likelihood-based scoring rules for comparing density forecasts in tails},
	volume = {163},
	year = {2011}}

@misc{brown74,
	author = {Brown, Thomas A.},
	howpublished = {Manuscript P-5235, RAND Corporation, Santa Monica, CA},
	title = {Admissible Scoring Systems for Continuous Distributions},
	year = {1974}}

@article{gneiting+r07,
	author = {Tilmann Gneiting and Adrian E Raftery},
	journal = {Journal of the American Statistical Association},
	number = {477},
	pages = {359-378},
	title = {Strictly Proper Scoring Rules, Prediction, and Estimation},
	volume = {102},
	year = {2007}}

@article{good52,
	author = {Good, I. J.},
	journal = {Journal of the Royal Statistical Society: Series B},
	number = {1},
	pages = {107-114},
	title = {Rational Decisions},
	volume = {14},
	year = {1952}}

@article{beaumont03,
	author = {Beaumont, Mark A},
	journal = {Genetics},
	number = {3},
	pages = {1139-1160},
	title = {Estimation of Population Growth or Decline in Genetically Monitored Populations},
	volume = {164},
	year = {2003}}

@article{andrieu+r09,
	author = {Christophe Andrieu and Gareth O. Roberts},
	journal = {The Annals of Statistics},
	number = {2},
	pages = {697 -- 725},
	title = {{The pseudo-marginal approach for efficient Monte Carlo computations}},
	volume = {37},
	year = {2009}}

@article{andrieu+dh10,
	author = {Andrieu, Christophe and Doucet, Arnaud and Holenstein, Roman},
	journal = {Journal of the Royal Statistical Society: Series B (Statistical Methodology)},
	number = {3},
	pages = {269-342},
	title = {Particle {M}arkov chain {M}onte {C}arlo methods},
	volume = {72},
	year = {2010}}

@inproceedings{lacoste-julian+hg11,
	author = {Lacoste--Julien, Simon and Husz{\'a}r, Ferenc and Ghahramani, Zoubin},
	booktitle = {Proceedings of the Fourteenth International Conference on Artificial Intelligence and Statistics},
	editor = {Gordon, Geoffrey and Dunson, David and Dud{\'\i}k, Miroslav},
	pages = {416--424},
	publisher = {PMLR},
	series = {Proceedings of Machine Learning Research},
	title = {Approximate inference for the loss-calibrated {B}ayesian},
	volume = {15},
	year = {2011}}

@article{picchini+t23,
    author = {Umberto Picchini and Massimiliano Tamborrino},
	journal = {Bayesian Analysis},
    volume = {20},
	number = {4},
	pages = {1283--1314},
	title = {Guided sequential {ABC} schemes for intractable {B}ayesian models},
	year = {2025}}

@incollection{fan+s18,
	address = {Boca Raton, Florida},
	author = {Fan, Y. and Sisson, S.A.},
	booktitle = {{Handbook of Approximate Bayesian Computation}},
	editor = {Sisson, S.A. and Fan, Y. and Beaumont, M.},
	pages = {87 -- 123},
	publisher = {CRC Press, Taylor \& Francis Group},
	series = {Chapman \& Hall/CRC Handbooks of Modern Statistical Methods},
	title = {{ABC} samplers},
	year = {2018}}

@article{loaiza-maya+mf21,
	author = {Loaiza-Maya, Ruben and Martin, Gael M. and Frazier, David T.},
	journal = {Journal of Applied Econometrics},
	number = {5},
	pages = {517-543},
	title = {Focused {B}ayesian prediction},
	volume = {36},
	year = {2021}}

@article{frazier+mmm19,
	author = {David T. Frazier and Worapree Maneesoonthorn and Gael M. Martin and Brendan P.M. McCabe},
	journal = {International Journal of Forecasting},
	number = {2},
	pages = {521-539},
	title = {Approximate {B}ayesian forecasting},
	volume = {35},
	year = {2019}}

\end{document}